\documentclass[amsmath,amssymb,nofootinbib,11pt,twocolumn,aps,letterpaper,superscriptaddress,showkeys,times,eqsecnum]{revtex4-2}

\usepackage{graphicx,epsfig}
\usepackage{amsmath}
\usepackage{amsfonts}
\usepackage{epsf}
\usepackage{epstopdf}
\usepackage {amssymb}
\usepackage{array}
\usepackage[linktocpage]{hyperref}
\usepackage[utf8]{inputenc}
\usepackage[usenames,dvipsnames]{color} 
\usepackage{hyperref}   
\usepackage[left=1.5cm,right=1.5cm,top=2.5cm,bottom=2.5cm]{geometry}
\usepackage{units}
\definecolor{oxfordblue}{rgb}{0.0, 0.13, 0.28}
\definecolor{burgundy}{rgb}{0.5, 0.0, 0.13}
\definecolor{darkolivegreen}{rgb}{0.33, 0.42, 0.18}
\definecolor{darkblue}{rgb}{0,0,0.7}
\definecolor{richcarmine}{rgb}{0.84, 0.0, 0.25}
\definecolor{bluer}{rgb}{0.00,0.50,0.75}{}
\hypersetup{colorlinks=true, citecolor=darkblue, linkcolor=richcarmine,
	urlcolor = bluer, filecolor=magenta}

\def\Ga{\Gamma}

\def\pt#1{\phantom{#1}}

\def\3g#1#2#3{^{(3)}\Ga^{#1}_{\pt{#1}#2#3}}

\newcommand{\beq}{\begin{equation}}
	\newcommand{\eeq}{\end{equation}}
\newcommand{\bea}{\begin{eqnarray}}
	\newcommand{\eea}{\end{eqnarray}}

\newcommand\bseq{\begin{subequations}} 
	\newcommand\eseq{\end{subequations}}
\newcommand\bcas{\begin{cases}}
	\newcommand\ecas{\end{cases}}

\begin{document}
	
\title{Hubble tension as a guide for refining the early Universe: Cosmologies with explicit local Lorentz and diffeomorphism violation}
	
	
	\author{Mohsen Khodadi}
	\email{m.khodadi@hafez.shirazu.ac.ir}
	\affiliation{Physics Department, College of Sciences, Shiraz University, Shiraz 71454, Iran}
	\affiliation{Biruni Observatory, College of Sciences, Shiraz University, Shiraz 71454, Iran}
	
	\author{Marco Schreck}
	\email{marco.schreck@ufma.br}
	\affiliation{Departamento de F\'{\i}sica, Universidade Federal do Maranh\~{a}o, \\
		Campus Universit\'{a}rio do Bacanga, S\~{a}o Lu\'{\i}s (MA), 65080-805, Brazil}
	\date{\today}
	
	\begin{abstract}
		
	This paper is dedicated to assessing modified cosmological settings based on the gravitational Standard-Model Extension (SME). Our analysis rests upon the Hubble tension (HT), which is a discrepancy between the observational determination of the Hubble parameter via data from the Cosmic Microwave Background (CMB) and Type Ia supernovae, respectively. While the latter approach is model-independent, the former highly depends on the model used to describe the physics of the early Universe. Motivated by the HT, we take into account two recently introduced cosmological models as test frameworks of the pre-CMB era. These settings involve local Lorentz and diffeomorphism violation parameterized by nondynamical SME background fields $s_{00}$ and $s^{ij}$, respectively. We aim at explaining the tension in the measured results of the cosmic expansion rate in early and late epochs by resorting to these two modified cosmologies as potential descriptions of the pre-CMB era. As long as the HT does not turn out to be a merely systematic effect, it can serve as a criterion for exploring regions of the parameter space in certain pre-CMB new-physics candidates such as SME cosmologies. By setting extracted limits on SME coefficients into perspective with already existing bounds in the literature, we infer that none of the aforementioned models are suitable pre-CMB candidates for fixing the HT. In this way, new physics arising from the particular realizations of Lorentz and diffeomorphism violation studied in this article does not explain the HT. Our paper exemplifies how to exploit this discrepancy as a novel possibility of refining our description of the early Universe.
				
	\end{abstract}
	
	
\maketitle
	
\section{Introduction}

The era of precision cosmology, in essence, began with the first measurements made by WMAP \cite{WMAP:2003elm,WMAP:2006jqi,WMAP:2008lyn}. Since then, additional terrestrial and orbital-based telescopes have started operating, which has allowed us to confront the standard model of cosmology, $\Lambda$-Cold Dark Matter ($\Lambda$CDM), with data. This brings us into a position to test the time evolution of our Universe, which the scientific community benefits from in two different manners. First, we can measure the $\Lambda$CDM parameters incredibly precisely. Second, because of this, we are in a position to investigate whether tiny deviations from standard cosmology might occur. Apart from the advantages that the advent of acquiring high-precision data introduced into cosmology, one of its inevitable side effects is the emergence of tensions.

In this regard, one of the most significant present discrepancies is the \textit{Hubble tension} (HT), which addresses
a mismatch in the current value of the Hubble parameter $H_0$ obtained by the Planck collaboration and the Hubble Space Telescope group, respectively. The first bases their analysis on data of the \textit{Cosmic Microwave Background} (CMB), which leads to an estimate of $H_{\mathrm{CMB}} = \unit[67.40 \pm 0.50]{km\,s^{-1}Mpc^{-1}}$~\cite{Planck:2018vyg}. The second employs \textit{SNeIa data}, which provides $H_{\mathrm{SNeIa}} = \unit[74.03 \pm 1.42]{km\,s^{-1}Mpc^{-1}}$~\cite{Riess:2019cxk}. In other words, it has been argued that the value attributed to the Hubble parameter in the early Universe reveals a meaningful deviation at the $4.4\,\sigma$ level from a wide set of late-time measurements; see Refs.~\cite{Freedman:2019jwv,Wong:2019kwg,Yuan:2019npk,LIGOScientific:2019zcs,Pesce:2020xfe,Soltis:2020gpl,Freedman:2021ahq} for additional details. Note that the mentioned deviation may even turn more significant, depending on the choice of data combinations~\cite{Lin:2019htv}. Like supernovae, lensing time delays \cite{Bonvin:2016crt,Birrer:2018vtm} through local measurements also suggest a higher value for the Hubble parameter in comparison to that inferred from the CMB.

Although there are many theoretical explanations that try to fix this discrepancy, the nature of its mechanism has not been disclosed, yet. In total, the proposals to account for the HT typically refer to either unknown systematic errors in the data or new physics beyond both $\Lambda$CDM and the Standard Model (SM) of elementary particle physics. It seems unlikely that this problem is resolved by introducing new particles or interactions \cite{Bernal:2016gxb,Feeney:2017sgx,Aylor:2018drw,Ye:2021nej}.
In the context of cosmology, novel proposals have been made recently to remedy the issue \cite{Vagnozzi:2019ezj,Barker:2020gcp,DiValentino:2021izs}, which also address the origin of the HT via a vacuum phase transition \cite{DiValentino:2017rcr}, Lifshitz vacuum energy \cite{Berechya:2020vcy}, early Dark Energy \cite{Poulin:2018cxd,Haridasu:2020pms,Kamionkowski:2022pkx}, early Dark Energy plus other components \cite{Ye:2021iwa,Reeves:2022aoi}, interacting Dark Energy \cite{DiValentino:2017iww,DiValentino:2019ffd,DiValentino:2019jae}, negative Dark Energy density \cite{Visinelli:2019qqu}, early recombination \cite{Sekiguchi:2020teg,Schoneberg:2021qvd},
the Dark Matter neutrino interaction \cite{DiValentino:2017oaw},
self-interacting neutrinos \cite{Kreisch:2019yzn,RoyChoudhury:2020dmd,Brinckmann:2020bcn}, and other effects~\cite{Mortsell:2018mfj,Yang:2018qmz,Aloni:2021eaq,Schoneberg:2022grr,Moshafi:2022mva} (see also review paper \cite{Abdalla:2022yfr}).

By resorting to the standard Heisenberg uncertainty principle at cosmological scales \cite{Capozziello:2020nyq,Spallicci:2021kye} as well as some of its generalized versions as semi-classical approaches to quantum gravity (QG)~\cite{Aghababaei:2021gxe}, further interesting ideas for solving the HT have been brought forward, too. Taking QG into account can be relevant, as quantum fluctuations were produced during the Planck epoch, which, after propagating through spacetime, generated primordial fluctuations in the inflationary era. In this way, the detection of anisotropies in the CMB may indicate imprints of QG \cite{Cai:2014hja,Kempf:2018gbt,Calmet:2019tur}, since the primordial fluctuations are quantified by a power spectrum \cite{Abazajian:2013vfg,Kaya:2021oih}. Some studies such as \cite{Tsujikawa:2003gh,Koivisto:2010fk,Krauss:2013pha,Ashtekar:2020gec} lead to the expectation that the quantum regime of gravity be encoded in the CMB.

Besides, measurements of the Hubble parameter based on the primordial Universe are vastly model-dependent, meaning that one cannot infer its value from the early Universe without already assuming a cosmological model \cite{Bernal:2016gxb}. This feature holds for analyses based on CMB data as well as measurements related to characteristics of the early Universe such as baryon acoustic oscillations. The situation is different for measurements that rely on SNeIa, which are, in essence, model-independent ones. Thus, taking into account the physics of the pre-CMB era including any modification of early-time cosmology can potentially be a well-motivated project to understand the root cause of the HT \cite{Vagnozzi:2021gjh}.
It has also been argued that utilizing the ages of high-redshift old astrophysical objects in the context of cosmology (without imposing assumptions on the pre-CMB expansion history), has the potential capability of alleviating the HT \cite{Vagnozzi:2021tjv}.

Another source of new physics, which could have an impact on cosmology, are violations of spacetime symmetries. After a spontaneous violation of Lorentz invariance had been shown to potentially occur due to fundamental physics at the Planck scale such as strings~\cite{Kostelecky:1988zi,Kostelecky:1989jp,Kostelecky:1989jw,Kostelecky:1991ak,Kostelecky:1994rn}, Colladay and Kosteleck\'{y} constructed the Standard-Model Extension (SME)~\cite{Colladay:1996iz,Colladay:1998fq} to furnish an effective, comprehensive description of Lorentz violation in field theory. The SME first emerged as an extension of the particle sectors in the SM. It parameterizes Lorentz violation in terms of background fields that couple to the SM fields such that the resulting contributions are coordinate-invariant. The background fields give rise to preferred spacetime directions where the size of Lorentz violation is governed by the controlling coefficients.

Apart from strings, mechanisms that result in a breakdown of Lorentz symmetry at the fundamental level were also identified in loop quantum gravity~\cite{Gambini:1998it,Bojowald:2004bb}, theories of noncommutative spacetimes~\cite{AmelinoCamelia:1999pm,Carroll:2001ws,Bailey:2018ifc}, spacetime foam~\cite{Klinkhamer:2003ec,Bernadotte:2006ya,Hossenfelder:2014hha}, nontrivial spacetime topologies~\cite{Klinkhamer:1998fa,Klinkhamer:1999zh,Klinkhamer:2002mj,Ghosh:2017iat}, Einstein-aether gravity \cite{Jacobson:2000xp,Jacobson:2007fh,Jacobson:2007veq,Khodadi:2020gns}, bumblebee gravity \cite{BG2005Kost,BG2005Bert,Seifert:2009gi,Maluf:2014dpa,Casana:2017jkc,Maluf:2020kgf,Maluf:2021lwh,Poulis:2021nqh,Khodadi:2022dff,Khodadi:2022mzt}, Ho\v{r}ava gravity~\cite{Horava:2009uw}, and covariant Ho\v{r}ava-like gravity \cite{Nojiri:2009th,Nojiri:2010kx,Cognola:2016gjy}.
Besides, in Refs.~\cite{Lambiase:2017adh,Gomes:2022hva} links have been established between certain SME coefficients and deformation parameters employed in the formulation of a generalized uncertainty principle. These results provide additional motivation for exploring the extent to which spacetime symmetries are exact.

Only few years after publishing the seminal papers~\cite{Colladay:1996iz,Colladay:1998fq}, the interest of the community turned towards conceiving an extension of General Relativity (GR) that is invariant under general coordinate transformations, but noninvariant under diffeomorphisms and/or local Lorentz transformations~\cite{Kostelecky:2003fs}. Diffeomorphism violation occurs in the presence of nondynamical background fields depending on the spacetime coordinates, whereas local Lorentz violation is implied by homogeneous background fields in freely falling inertial frames. In the context of explicit symmetry violation, both phenomena are not directly linked to each other, but must be considered as different manifestations of physics beyond GR. The recent papers \cite{Kostelecky:2020hbb,Kostelecky:2021tdf} extensively complement the very first work \cite{Kostelecky:2003fs}, which introduced the gravitational sector of the SME more than 15 years ago.

Since then, the gravitational SME has found various applications in modified-gravity phenomenology \cite{Bailey:2006fd,Bailey:2009me,Tso:2011up,Bailey:2013oda,Shao:2014oha,Shao:2014bfa,Long:2014swa,Kostelecky:2015dpa,Shao:2016cjk,Bonder:2017dpb,Bonder:2020fpn,ONeal-Ault:2021uwu,Bailey:2022wuv,Haegel:2022ymk}. In parallel with the phenomenological articles, purely theoretical work has been done such as in Refs.~\cite{Bluhm:2007bd,Bluhm:2014oua,Bonder:2015maa,Bluhm:2016dzm,Bluhm:2019ato,Bluhm:2021lzf,ONeal-Ault:2020ebv,Reyes:2021cpx,Nilsson:2022mzq,Reyes:2022mvm,Reyes:2022ipv,Reyes:2022dil} improving our understanding of symmetry violation in gravity. It is also worthwhile to mention the review paper \cite{Hees:2016lyw} which, within the framework of the gravitational SME, presents a compilation of well-known tests of Lorentz symmetry violation. In this respect, Ref.~\cite{Khodadi:2022pqh} can also be consulted for a brief list of references along with a novel exploration of Lorentz symmetry violation at the black-hole horizon scale.

Cosmological evolution based on different sectors of the gravitational SME has been on the focus in several very recent articles \cite{Bonder:2017dpb,ONeal-Ault:2020ebv,Nilsson:2022mzq,Reyes:2022dil}, which has ignited the emergence of a brand-new research direction. In the present work, this promising area is to be applied to the explanation of certain discrepancies in experimental data that cannot be accounted for in the context of $\Lambda$CDM. One of the key concepts in GR, which according to eminent QG candidates may be violated in the pre-CMB period, are diffeomorphism invariance and local Lorentz invariance. According to the previous discussion, the CMB spectrum can be prone to carrying signals for violations of the aforementioned symmetries. Hence, the gravitational SME is more than suitable as a test framework in a cosmological setting.

In this regard, we take two cosmological scenarios~\cite{ONeal-Ault:2020ebv,Reyes:2022dil} into account, which have recently been inspired by the gravitational SME. Given that results inferred from CMB data strongly depend on early-Universe physics, our objective is to investigate whether or not these two modified cosmologies related to the high-energy phase of the Universe are capable of explaining the HT. More exactly, assuming that the HT is not simply caused by systematics, any potential cosmological model related to the pre-CMB era should be able to provide a compatible explanation for it. In this way, we are dealing with a novel possibility of refining high-energy cosmologies beyond $\Lambda$CDM.

Our manuscript is organized as follows. Section~\ref{sec:cos} shall provide a brief review of the two models that we intend to analyze in this paper. Section~\ref{sec:shedding-light-HT} is dedicated to the problem of the HT that is considered from the perspective of the models previously referred to. Section~\ref{sec:conclusions} contains a brief summary of our findings. Throughout this paper natural units will be used with $c=1$ unless otherwise stated. Greek indices are reserved for quantities living in four-dimensional spacetime manifolds, whereas Latin indices are employed for purely spacelike parts of tensors. Latin indices with a bar on top are used in local inertial frames.

\section{Diffeomorphism-violating cosmological solutions}
\label{sec:cos}

In this section, we shall briefly review cosmological solutions based on the SME, which have recently been derived in Refs.~\cite{ONeal-Ault:2020ebv,Reyes:2022dil}. Actually, these solutions, which involve nondynamical SME coefficients denoted as $u$, $s_{\mu\nu}$, and $s^{\mu\nu}$ in the literature, address the role of local Lorentz and diffeomorphism symmetry violation in cosmological time evolution.

References~\cite{ONeal-Ault:2020ebv,Reyes:2022dil} both benefit from the $(3+1)$ decomposition of gravity, which is also known as the ADM formalism \cite{Arnowitt:1962hi,Arnowitt:2008}. The latter is a very useful tool that allows us to represent a four-dimensional spacetime manifold as a continuous family of spacelike hypersurfaces $\Sigma_t$ evolving in time. One implication of this procedure is that the quantities describing curvature also decompose into contributions completely defined in $\Sigma_t$, quantities living in subspaces orthogonal to $\Sigma_t$, and mixed contributions. Similarly, such decompositions can be performed for SME background fields. Working in the ADM formalism has proven quite fruitful and renders some properties of a modified-gravity theory more transparent than does the covariant approach; see Refs.~\cite{ONeal-Ault:2020ebv,Reyes:2021cpx,Reyes:2022dil,Reyes:2022mvm,Reyes:2022ipv} for applications of this formalism to the SME.

\subsection{First scenario}
\label{sec:first-scenario}

The authors of Ref.~\cite{ONeal-Ault:2020ebv} deal with the following modified Einstein-Hilbert (EH) action inspired from Ref.~\cite{Kostelecky:2003fs} (see also Ref.~\cite{Kostelecky:2020hbb}) in the presence of a cosmological constant $\Lambda$:
\begin{subequations}
	\label{lag1}
	\begin{align}
		S&=\int_{\mathcal{M}} \mathrm{d}^4x\,(\mathcal{L}^{(0)}+\mathcal{L}_{\mathrm{SME}})+S_m\,, \displaybreak[0]\\[2ex]
		\mathcal{L}^{(0)}&=\frac{\sqrt{-g}}{2\kappa}({}^{(4)}R-2\Lambda)\,, \displaybreak[0]\\[2ex]
		\mathcal{L}_{\mathrm{SME}}&=\frac{\sqrt{-g}}{2\kappa}\Big[-u{}^{(4)}R + s_{\mu \nu}({}^{(4)}R^{(T)})^{\mu\nu} \notag \\
		&\phantom{{}={}}\hspace{1.1cm}+t_{\mu\nu\rho\sigma}{}^{(4)}C^{\mu\nu\rho\sigma}\Big]\,,
	\end{align}
\end{subequations}
with $\kappa\equiv8\pi G_N$ and where $g\equiv\det(g_{\mu\nu})$ is the determinant of the metric $g_{\mu\nu}$ of the four-dimensional spacetime manifold $\mathcal{M}$. In the action above, the geometrical quantities ${}^{(4)}R\equiv {}^{(4)}R^{\mu}_{\phantom{\mu}\mu}$, $({}^{(4)}R^{(T)})^{\mu\nu}$, and ${}^{(4)}C^{\mu\nu\rho\sigma}$ are the Ricci scalar, the traceless Ricci tensor, and the Weyl curvature tensor, respectively, on $\mathcal{M}$. The coefficient $u$ is a scalar nondynamical background field, whereas $s_{\mu\nu}$ and $t_{\mu\nu\rho\sigma}$ denote tensor-valued ones. By field redefinitions, the coefficients $u$ and $s_{\mu\nu}$ can be transferred into the matter sector described by the action $S_m$ \cite{Bonder:2015maa,Bluhm:2019ato}. Conventions are adopted such that the SME coefficients reside in the gravity sector.

Concerning the modified action of Eq.~\eqref{lag1}, several points should be highlighted. First, it is invariant under general coordinate transformations, as all covariant objects are properly contracted. However, it breaks diffeomorphism symmetry, since --- unlike the metric and other dynamical fields --- the background fields $u$, $s_{\mu \nu}$, and $t_{\mu \nu \rho \sigma}$ do not transform as tensors and, in essence, stay fixed. It is possible to define new background fields in a freely falling inertial frame from the background fields $s_{\mu\nu}$ and $t_{\mu\nu\varrho\sigma}$ via the vierbein formalism~\cite{Kostelecky:2020hbb}. The latter imply a violation local Lorentz invariance, since these newly defined background fields stay fixed under local Lorentz transformations carried out at any spacetime point. It is also worthwhile to emphasize that in the modified action of Eq.~(\ref{lag1}), homogeneity is absent, as the background fields in a curved spacetime manifold are necessarily coordinate-dependent \cite{Kostelecky:2003fs}.

Generally, the presence of $s_{\mu\nu}$ in Eq.~\eqref{lag1} breaks diffeomorphism symmetry explicitly, since the background field is put into the EH action by hand and it is not subject to an action principle. Local Lorentz invariance explicitly breaks down with a nondynamical background being defined in local inertial frames, whereby a breakdown of spatial isotropy can be a consequence of the latter. In contrast, if the background fields are described to have their own dynamics, diffeomorphism and local Lorentz breaking occur spontaneously implying different physics; see Refs.~\cite{Bluhm:2019ato,Bluhm:2016dzm} for detailed discussions.

Based on the modified action stated in Eq.~\eqref{lag1}, the modified Einstein equations read
\begin{equation}
	G^{\mu\nu} = (T^{ust})^{\mu\nu} + \kappa (T_m)^{\mu\nu}\,,
	\label{SMEfe}
\end{equation}
where $G^{\mu\nu}=R^{\mu\nu}-(R/2)g^{\mu\nu}$ is the Einstein tensor and $(T_m)^{\mu\nu}$ the energy-momentum tensor related to the standard matter sector. Furthermore, $(T^{ust})^{\mu\nu}$ involves the coefficients $u$, $s_{\mu \nu}$, and $t_{\mu \nu \rho \sigma}$ where its explicit form can be found in Ref.~\cite{Kostelecky:2003fs}. The second Bianchi identities of pseudo-Riemannian geometry, $\nabla_\mu G^{\mu\nu} =0$, require that
\begin{equation}
	\nabla_\mu (T^{ust})^{\mu\nu} =- \kappa \nabla_\mu (T_m)^{\mu\nu}\,,
	\label{SMEcl}
\end{equation}
which correspond to four conservation laws that must be satisfied by hand whenever symmetry breaking is explicit. Note that for spontaneous symmetry breaking, these Noether identities are satisfied automatically on-shell.

In what follows, the fourth-rank tensor background $t_{\mu\nu\varrho\sigma}$ is discarded, since its treatment complicates the analysis to a large degree (see Refs.~\cite{Bonder:2015maa,Bonder:2020fpn}).
Moreover, the forthcoming study rests upon the \textit{ansatz} of a Friedmann-Lema\^{i}tre-Robertson-Walker (FLRW) spacetime with metric
\begin{subequations}
	\label{flrw}
	\begin{align}
		\mathrm{d}s^2&= - \mathrm{d}t^2 +h_{ij}\mathrm{d}x^i\mathrm{d}x^j\,,\quad h_{ij}=a^2(t) \tilde{g}_{ij}\,, \\[2ex]
		\tilde{g}_{ij}&=\left(\frac{1}{1-kr^2},r^2,r^2\sin\theta\right)\,,
	\end{align}
\end{subequations}
where $a(t)$ is the time-dependent scale factor and spherical coordinates $(r,\theta,\phi)$ are used in the purely spacelike part $\tilde{g}_{ij}$. In a suitable set of coordinates, the curvature scalar $k=\{+1,0,-1\}$ describes a closed, flat, and open Universe, respectively.

The FLRW metric relies on homogeneity and spatial isotropy of the underlying spacetime manifold. In Ref.~\cite{Reyes:2022dil} we argued that $|s_{\mu\nu}|\ll 1$ should hold such that the \textit{ansatz}~(\ref{flrw}) can be considered a reasonable choice for studying the cosmological implications of the modified-gravity theories consulted here. In particular, for coefficients in local inertial frames that imply a loss of spatial isotropy, the use of the FLRW metric as an \textit{ansatz} is still warranted as long as diffeomorphism- and Lorentz-violating effects are small. According to present bounds on components of $s_{\mu\nu}$ in the Sun-centered equatorial reference frame provided by the data tables~\cite{Kostelecky:2008ts}, this requirement is satisfied, i.e., it is safe to assume that $|s_{\mu\nu}|\ll 1$.

Combinations of coefficients that are exempt from this requirement are those independent of the spacetime coordinates, i.e., $\partial_{\varrho}s_{\mu\nu}=0$ in Cartesian coordinates, as they maintain homogeneity. Also, we refer to coefficients in local inertial frames defined by $s_{\overline{a}\overline{b}}\equiv e^{\mu}_{\phantom{\mu}\overline{a}}e^{\nu}_{\phantom{\nu}\overline{b}}s_{\mu\nu}$ with local-frame indices $\overline{a},\overline{b}\in \{\overline{0},\overline{1},\overline{2},\overline{3}\}$ and background vierbeins satisfying $\eta_{\overline{a}\overline{b}}=e^{\mu}_{\phantom{\mu}\overline{a}}e^{\nu}_{\phantom{\mu}\overline{b}}g_{\mu\nu}$ with the Minkowski metric $\eta_{\overline{a}\overline{b}}$. Then, sectors with a nonzero purely timlike coefficient $s_{\overline{0}\overline{0}}$ or nonzero diagonal, purely spacelike coefficients chosen as equal, $s_{\overline{1}\overline{1}}=s_{\overline{2}\overline{2}}=s_{\overline{3}\overline{3}}$, do not break local isotropy. Thereupon, these configurations are not in contradiction with the FLRW metric. Such sectors have been studied in the literature, e.g., in Ref.~\cite{ONeal-Ault:2020ebv}. However, the SME coefficients in the models considered, e.g., Eq.~\eqref{lag1} are not given in local inertial frames, but in generic spacetime frames. Note also that the authors of Ref.~\cite{ONeal-Ault:2020ebv}, for simplicity, assumed that $s_{00}$ is
independent of the spatial coordinates.

Applying Hamilton's equations to a sector of Eq.~\eqref{lag1} with one nonzero component $s_{00}$ of $s_{\mu\nu}$ leads to the following dynamical equations:
\begin{subequations}
	\label{eq:modified-friedmann-model-1}
	\begin{align}
		\label{f1}
		(1-s_{00})\left(\frac{\dot{a}}{a}\right)^2&=\frac {\kappa }{3}\rho - \frac{k}{a^2} \notag \\
		&\phantom{{}={}}-s_{00} \frac{\ddot{a}}{a} + \frac {\dot{a}}{a} \frac {\dot{s}_{00}}{2}\,, \\[2ex]
		\label{f2}
		(1-s_{00})\left[\frac {\ddot{a}}{a} + \frac 12 \left(\frac{\dot{a}}{a}\right)^2 \right]
		&= -\frac {\kappa}{2}p - \frac{k}{2a^2} \notag \\
		&\phantom{{}={}}+ \frac {\dot{a}}{a} \dot{s}_{00} + \frac{\ddot{s}_{00}}{4}\,,
	\end{align}
\end{subequations}
where a dot stands for a time derivative. Recall that matter is assumed to be standard. Therefore, the latter modified Friedmann equations rest upon the standard perfect-fluid model for a homogeneous and isotropic Universe, i.e., $(T_m)^\mu_{\phantom{\mu}\nu} = {\rm diag}(-\rho, p,p,p)$. Here, $\rho$ and $p$ represent the matter energy density and pressure, respectively, which are related via the equation of state $p= w \rho$ with the barotropic index $w$. The modifications of the Friedmann equations \eqref{eq:modified-friedmann-model-1} include terms with first- and second-order time derivatives of $s_{00}$, scalings by $1-s_{00}$, and also an extra $\ddot{a}$ term in the first equation.

It is clear that $s_{00}\mapsto 0$ recovers the standard Friedmann equations, which makes sense, as this limit switches off explicit diffeomorphism symmetry breaking in the action of Eq.~\eqref{lag1}. For a static $s_{00}$, the spatial components of Eq.~\eqref{SMEcl} are automatically satisfied. However, the timelike component gives rise to an extra requirement. So for $\nu=0$ the left-hand side of Eq.~\eqref{SMEcl} reads~\cite{ONeal-Ault:2020ebv}
\begin{subequations}
	\begin{equation}
		\nabla_\mu (T^s)^\mu_{\phantom{\mu} 0}=   \frac {\ddot{a}}{a} \left( \frac{3}{2}\dot{s}_{00} + 6 s_{00} \frac{\dot{a}}{a} \right)+3 s_{00} \frac {\dddot{a} }{a}\,,
		\label{FLRWcons0}
	\end{equation}
	where $(T^s)^{\mu\nu}$ corresponds to $(T^{ust})^{\mu\nu}$ of Ref.~\cite{Kostelecky:2003fs} with $u=t^{\mu\nu\varrho\sigma}=0$. The right-hand side of Eq.~\eqref{SMEcl}, which refers to the matter sector, involves
	\begin{equation}
		\nabla_\mu (T_m)^\mu_{\phantom{\mu} 0}= -\dot{\rho} - 3\frac{\dot{a}}{a}\left(\rho+p\right)\,.
		\label{FLRWconsMatter0}
	\end{equation}
\end{subequations}
The consistency conditions of Eq.~\eqref{SMEcl} allow for deriving the form of $s_{00}$ by solving for $\rho$ and $p$ in the modified Friedmann equations~\eqref{eq:modified-friedmann-model-1} and inserting the relevant expressions into Eq.~\eqref{FLRWconsMatter0}. Naturally, different choices of $s_{00}$ result in a different cosmological dynamics according to the modified Friedmann equations~\eqref{eq:modified-friedmann-model-1}. As mentioned before, the background field is assumed to be nondynamical, i.e., $s_{00}$ is a prescribed quantity chosen by hand.

However, Eq.~\eqref{SMEcl} is mandatory and can serve as a relationship that restricts $s_{00}$. Therefore, the authors of Ref.~\cite{ONeal-Ault:2020ebv} take the reasonable decision to look at two particular ways to satisfy Eq.~\eqref{SMEcl}. For the first case they demand that the matter stress-energy tensor by itself be completely conserved and thereby, the expressions of Eqs.~(\ref{FLRWcons0}), (\ref{FLRWconsMatter0}) must vanish independently of each other. For the second case they require that the total conservation law of Eq.~\eqref{SMEcl} hold, whereas the matter stress-energy tensor is not necessarily conserved separately.

Starting with the first possibility, i.e., enforcing energy-momentum conservation for matter, it is necessary to deal with $\nabla_\mu (T^s)^\mu_{\phantom{\mu} 0} = 0$, which results in the following analytical solution for $s_{00}$:
\begin{equation}
	s_{00} = \frac {\zeta}{a^4 \ddot{a}^2 }\,,
	\label{s00soln}
\end{equation}
where $\zeta$ is an arbitrary constant. The solution above has pathological features, e.g., it diverges when the acceleration vanishes, $\ddot{a}=0$.
To crosscheck whether this solution can be considered a physical one, it is inserted into the modified Friedmann equations \eqref{eq:modified-friedmann-model-1} to solve the resulting system of equations for $a(t)$ with different choices of sources $\rho$ and $p$. However, these equations contain up to fourth-order time derivatives and it is impractical to determine an exact analytical solution. A challenge is posed even by adopting a perturbative point of view, meaning that again when $\ddot{a}$ approaches zero, the dimensionless $s_{00}$ becomes large, which is in conflict with perturbation theory. Therefore, the first choice of $s_{00}$ results in an inappropriate cosmological solution and will be discarded.

As for the second choice, the authors of Ref.~\cite{ONeal-Ault:2020ebv} do not require that $\nabla_\mu (T^s)^\mu_{\phantom{\mu} 0}=0$. Then, $s_{00}$ is not restricted in any way \textit{a priori} so that the simplest case to consider is that of a static $s_{00}$, i.e., $\dot{s}_{00}=0$. As a result of this choice, the modified Friedmann equations in the absence of the cosmological constant $\Lambda$ take the following form:
\begin{subequations}
	\label{eq:friedmann-equations-model1}
	\begin{align}
		H^2 &= \frac {\kappa \rho}{3 (1-3s_{00}/2 )} - \frac {k}{ a^2 (1-s_{00}) } \notag \\
		&\phantom{{}={}}+ \frac {\kappa p s_{00} }{(2-3 s_{00})(1-s_{00})}\,,
		\label{feqns:set21} \displaybreak[0]\\[1ex]
		\dot{H}+H^2 &= - \frac {\kappa (\rho + 3 p)}{6 (1-3s_{00}/2 )}\,,
		\label{feqns:set2}
	\end{align}
\end{subequations}
where $H=H(t)=\dot{a}(t)/a(t)$ is the Hubble parameter. Although the latter equations contain the usual GR terms with simple scaling factors, a term of a completely different structure, which involves the pressure, emerges in Eq.~\eqref{feqns:set21}. By using Eqs.~\eqref{FLRWconsMatter0} and \eqref{feqns:set2}, the modified conservation law or continuity equation for matter is obtained to have the form
\begin{subequations}
	\begin{equation}
		\dot{\rho}+3\frac{\dot{a}}{a}f(w,s_{00})\rho = 0\,,
		\label{eq:cont}
	\end{equation}
	with the auxiliary function
	\begin{equation}
		f(w,s_{00}) = \frac{2(1 + w -s_{00})} {2 +s_{00}(3w -2)}\,,
		\label{f}
	\end{equation}
\end{subequations}
which reduces to the proper GR limit, $f=1+w$, for $s_{00} \mapsto 0$. Integrating the modified continuity equation implies
\begin{equation}
	\rho=\rho_0\left(\frac{a}{a_0}\right)^{-3 f(w,s_{00})}\,,
	\label{rhmod}
\end{equation}
where $a_0$ is the present value of the scale factor. When matter is assumed to consist entirely of dust, $w=0$, cosmological evolution takes its standard form $\rho \sim a^{-3}$ due to $f=1$, which holds even for $s_{00}\neq 0$. In contrast, for a radiation-dominated stage with $w=1/3$ or Dark Energy with $w=-1$, the evolution equation is modified.

\subsection{Second scenario}
\label{sec:second-scenario}

The modified EH action employed in Ref.~\cite{Reyes:2022dil}, which bears similarities to that of Ref.~\cite{ONeal-Ault:2020ebv}, was proposed originally in Ref.~\cite{Kostelecky:2003fs}; see also Ref.~\cite{Kostelecky:2020hbb}. In contrast to the background fields used in Ref.~\cite{ONeal-Ault:2020ebv}, the SME coefficients now have upper Lorentz indices, i.e.,
\begin{subequations}
	\label{eq:action-initial}
	\begin{align}
		S&=\int_{\mathcal{M}} \mathrm{d}^4x\,(\mathcal{L}^{(0)}+\mathcal{L}_{\mathrm{SME}})+S_m\,, \displaybreak[0]\\[2ex]
		\mathcal{L}^{(0)}&=\frac{\sqrt{-g}}{2\kappa}({}^{(4)}R-2\Lambda)\,, \displaybreak[0]\\[2ex]
		\mathcal{L}_{\mathrm{SME}}&=\frac{\sqrt{-g}}{2\kappa}(-u {}^{(4)}R + s^{\mu \nu} {}^{(4)}R^T_{\mu \nu}+t^{\mu \nu \rho \sigma} {}^{(4)}C_{\mu \nu \rho \sigma})\,,
	\end{align}
\end{subequations}
where the geometrical quantities are defined such as in Eq.~\eqref{lag1} and the SME background fields $s^{\mu\nu},t^{\mu\nu\varrho\sigma}$ share the same generic properties with those of Eq.~\eqref{lag1}. Again, the background field $t^{\mu\nu\varrho\sigma}$ is not taken into account. Then, the action of the model persued in Ref.~\cite{Reyes:2022dil} takes the form \begin{align}
	\label{eq:actionus}
	S &= \int_{\mathcal{M}} \mathrm{d}^4x\,\frac{\sqrt{-g}}{2\kappa}\left[(1 - u) {}^{(4)}R + s^{\mu \nu} {}^{(4)}R_{\mu\nu}-2\Lambda\right] \notag \\[1ex]
	&\phantom{{}={}}+S_m\,,
\end{align}
where in comparison with the action of Eq.~\eqref{eq:action-initial}, the trace of the Ricci tensor is kept for simplicity.

The following simplifying conditions will be imposed on the underlying modified-gravity theory defined by the action of Eq.~\eqref{eq:actionus}:
\begin{equation}
	\label{eq:consistency-conditions}
	\dot{u} =\dot{s}^{i j} =\dot{s}^{00} = 0\,,
\end{equation}
which render the SME coefficients static; cf.~the choice of $s_{00}$ taken in the paragraph above Eq.~\eqref{eq:friedmann-equations-model1} in the second case of the first scenario. These requirements lead to extensive computational simplifications. Then, for the FLRW spacetime with the metric of Eq.~\eqref{flrw} the dynamical field equations in the presence of the coefficients $u$ and $s^{\mu\nu}$ read
\begin{subequations}
	\label{eq:modified-friedmann-model-2}
	\begin{align}
		H^2&=\frac{1}{3(\Upsilon-s^{00})}\left(\rho+\Lambda-3\Upsilon\frac{k}{a^2}-D_iD^iu \right. \notag \\
		&\phantom{{}={}}\hspace{2.1cm}\left.{}-\frac{1}{2}D_iD^is^{00}+\frac{1}{2}D_iD_js^{ij}\right)\,,\label{eq:constraint1} \displaybreak[0]\\[2ex]
		\label{eq:secfriedmannb}
		\dot{H}+H^2&=-\frac{1}{6(\Upsilon-s^{00})}\!\left[\rho+3P-2\Lambda+2s\left(\frac{k}{a^2}+H^2\right)\right. \notag \\
		&\phantom{{}={}}\hspace{2.3cm}+D_iD^iu-\frac{1}{2}D_iD^is^{00} \notag \\
		&\phantom{{}={}}\hspace{2.3cm}\left.{}-\frac{1}{2}D_iD^is\right]\,,
	\end{align}
\end{subequations}
where $\Upsilon= 1 - u + s/3$, with the trace $s$ of the purely spacelike part $s^{ij}$, i.e., $s\equiv s^{ij}h_{ij}=a^2s^{ij}\tilde{g}_{ij}$. Here, the symbol $D_i$ represents a covariant derivative compatible with the induced metric on $\Sigma_t$. The modified Friedman equations involve additional contributions from the background fields $u$, $s^{00}$, and $s^{ij}$.
Besides, the treatment of Eq.~\eqref{eq:action-initial} within the ADM formalism leads to a further constraint on the background coefficients:
\begin{equation}
	\label{eq:constraint2}
	4D_i\Big\{H\Big[(1-u-s^{00})h^{ik}+s^{ik}\Big]\Big\}=0\,,
\end{equation}
which plays a very significant role in the final form of the modified Friedman equations and the subsequent phenomenological analysis. As before, the presence of a generic $s^{\mu\nu}$ leads to a loss of homogeneity as well as spatial isotropy, when an inferred background is defined in a local inertial frame by means of the vierbein formalism. Both background fields are independent entities when symmetry breaking is explicit.

Since GR has survived all experimental searches for modified-gravity effects, the dimensionless SME coefficients satisfy $|s^{\mu\nu}|\ll 1$; see the subsequent paragraph under Eq.~\eqref{flrw} and the data tables~\cite{Kostelecky:2008ts}. Since Lorentz violation in the matter sector is neglected, the standard continuity equation $\nabla_\mu (T_m)^{\mu\nu}=0$ with Eq.~\eqref{FLRWconsMatter0} is assumed to hold for matter to a justifiable degree and SME coefficients are only considered in the modified Friedmann equations. In other words, the matter sector of the modified action of Eq.~(\ref{eq:actionus}) is assumed to be sourced by a perfect fluid. Therefore, the following relation holds between the matter density and the scale factor (cf.~Eq.~\eqref{rhmod}):
\begin{equation}
	\label{eq:rhowitha}
	\rho=\rho_0\left(\frac{a}{a_0}\right)^{-3 (1+w)}\,.
\end{equation}
As can be seen from the modified Friedmann equations \eqref{eq:friedmann-equations-model1}, \eqref{eq:modified-friedmann-model-2}, within the framework of the gravitational SME, alterations of the dynamical equations have the potential to impact the time evolution of the Universe. In Ref.~\cite{Reyes:2022dil}, for the sake of simplicity, three different sectors of the background fields are considered: i) the scalar background $u$ ($s^{\mu\nu}=0$), ii) the purely timelike background $s^{00}$ ($u=0=s^{ij}$), and iii) the tensor-valued purely spacelike background $s^{ij}$ ($u=0=s^{00}$).

In Ref.~\cite{Reyes:2022dil} it is also shown that when Eq.~\eqref{SMEcl} is applied to the first two cases, each of the background fields $u$ and $s^{00}$ merely gives rise to constant scaling factors in the evolution equations, which do not have a physical effect on the time evolution of the Universe. Thus, if the background fields $u$ and $s^{00}$ are incorporated separately into the EH action, the resulting cosmology does not deviate from that based on GR. The third case of $s^{ij}$, which we will dedicate our attention to, will tell a different story.

By setting $u=s^{00}=0$, the constraint of Eq.~\eqref{eq:constraint2} requires that
\begin{equation}
	\label{eq:constraintsij}
	D_i s^{ij}=0\,.
\end{equation}
A rearrangement of the modified Friedman equations \eqref{eq:modified-friedmann-model-2} combined with Eq.~(\ref{eq:constraintsij}), which is considered in a spatially flat Universe with $k=0$ and in the absence of $\Lambda$, leads to
\begin{subequations}
	\label{eq:friedmann-spatial}
	\begin{align}
		\label{eq:friedmann-1-spatial}
		H^2 &= \frac{\kappa}{3(1+s/3)} \rho\,, \\[2ex]
		\dot{H}+H^2&=-\frac{\kappa}{6(1+s/3)} \notag \\
		&\phantom{{}={}}\times\left(\rho+3p-\frac{1}{2}D_iD^is+2sH^2\right)\,.
	\end{align}
\end{subequations}
The dynamical equations above result in a stage of accelerated expansion provided that either the first or the second of the following sets of conditions is fulfilled:
\begin{subequations}
	\begin{align}\label{eq:bound-s}
		\rho + 3 P &\gtrless \frac{1}{2} D_i D^i s - 2 s H^2\,, \\[2ex]
		s & \gtrless -3\,. \label{eq:bound-ss}
	\end{align}
\end{subequations}
We select a particular example for the background tensor field $s^{\mu\nu}$ with nonzero purely spacelike entries,
\begin{equation}
	\label{eq:sijnew}
	s^{\mu \nu} = -\alpha \left(\begin{array}{cccc}
		0 & 0 & 0 & 0 \\
		0 & 1 & 0 & 0 \\
		0 & 0 & r^{-2} & 0 \\
		0 & 0 & 0 & r^{-2} \sin^{-2} \theta \\
	\end{array}\right)\,,
\end{equation}
where $\alpha$ is a dimensionless real parameter, which is independent of the spacetime coordinates and controls the size of $s^{ij}$. Then, Eq.~ \eqref{eq:constraintsij} provides $s=-3\alpha a^2$. Now, the dynamical equations \eqref{eq:friedmann-spatial} are reformulated as follows:
\begin{subequations}
	\begin{align}
		\label{eq:frdmod}
	&	H^2 = \frac{\kappa}{3 (1 - \alpha a^2)} \rho\,, \\[2ex]
		\label{eq:frdmod2}
	&	\dot{H} + H^2 = - \frac{\kappa}{6 (1 - \alpha a^2)} (\rho + 3 p - 6 \alpha a^2H^2)\,.
	\end{align}
\end{subequations}
After arriving at this form of the modified Friedmann equations, similarly to GR, Eq.~\eqref{eq:frdmod2} is, in essence, a direct consequence of Eq.~\eqref{eq:frdmod} and energy-momentum conservation for standard matter, Eq.~\eqref{eq:rhowitha}. Therefore, taking Eq.~\eqref{eq:frdmod} into account is sufficient to study the dynamics of the Universe. By inserting $s=-3\alpha a^2$ into Eq.~\eqref{eq:bound-ss}, one deduces the following constraint on the scale factor describing a Universe subject to an accelerated expansion:
\begin{equation}
	\label{eq:infws2}
	a^2<\frac{1}{\alpha}\,.
\end{equation}
Hence, for a fixed value of $\alpha$, the scale factor should not exceed a certain value. In other words, the presence of the background field $s^{ij}$ given by Eq.~\eqref{eq:sijnew} in the EH action results in an accelerated expansion of the Universe that stops at a particular instant. By inserting the first Friedmann equation \eqref{eq:frdmod} into \eqref{eq:bound-s}, using the equation of state $p=w\rho$ for a perfect fluid, and taking Eq.~\eqref{eq:infws2} into account, we have
\begin{equation}
	1 + 3 w < \frac{2 \alpha a^2}{1 - \alpha a^2} \Leftrightarrow \frac{1 + 3 w}{3 (1 + w) \alpha} < a^2\,,
\end{equation}
where for dust and radiation with $w=0$ and $w=1/3$, respectively, the scale factor lies within either one of the following ranges:
\begin{subequations}
	\begin{align}\label{eq:range-scale-factor-dust}
		\frac{1}{3 \alpha} < a^2 < \frac{1}{\alpha}\,,
		\\[2ex]
		\label{eq:range-scale-factor-radiation}
		\frac{1}{2 \alpha} < a^2 < \frac{1}{\alpha}\,.
	\end{align}
\end{subequations}
A numerical solution of Eq.~\eqref{eq:frdmod} was computed in Ref.~\cite{Reyes:2022dil}, which shows that the time frame where Eq.~\eqref{eq:infws2} is fulfilled coincides with $\ddot{a}>0$. So there is a short period when an accelerated expansion occurs and at the end of this period the scale factor turns imaginary and becomes physically meaningless. The origin of this behavior comes from the requirement of satisfying Eq.~\eqref{eq:bound-s} in the absence of exotic matter to drive accelerated expansion. Unlike in the scenario of Sec.~\ref{sec:first-scenario}, one can explain an accelerated expansion of the Universe within a certain time interval without resorting to any type of exotic matter.

\section{Can the previous scenarios explain the HT?}
\label{sec:shedding-light-HT}

Having reviewed the essential findings of Refs.~\cite{ONeal-Ault:2020ebv,Reyes:2022dil} in a cosmological context, we would like to figure out whether these particular cosmologies based on the SME are able to account for the HT. In other words, with the assumption that the HT is not ruled out in future measurements of the Hubble parameter, we will employ it as a criterion to survey the modified-cosmology scenarios discussed in Secs.~\ref{sec:first-scenario}, \ref{sec:second-scenario} above.

We adopt the simple and reasonable point of view that via the first modified Friedmann equation these two settings at hand address the Hubble parameter measured by the Planck collaboration based on CMB data. In contrast, the Hubble parameter measured by the HST group through SNeIa data is assumed to be related to the first Friedmann equation of standard cosmology.

Let us start with the first scenario of Sec.~\ref{sec:first-scenario}. For a spatially constant $s_{00}$ we redefine the coupling constant via $\kappa\mapsto \kappa_{\mathrm{eff}}=\kappa/(1-3s_{00}/2)$, since experiment cannot distinguish between $\kappa$ and $\kappa_{\mathrm{eff}}$. By incorporating the radiation content of a perfect fluid with equation of state $p=\rho/3$ as well as $k=0$ into the first modified Friedmann equation of Eq.~\eqref{feqns:set21}, a simple algebraic calculation leads to
\begin{equation}
	H_{\mathrm{CMB}}^2 =H^2_{\mathrm{SNeIa}}\left(1+ \frac {s_{00}}{2 (1-s_{00} )}\right)\,,
	\label{H2}
\end{equation}
where $H^2_{\mathrm{SNeIa}}=H^2_{\mathrm{GR}}=(\kappa_{\mathrm{eff}}/3)\rho$.
The modification proportional to $s_{00}/(1-s_{00})$ has to be negative, because of $H_{\mathrm{CMB}}<H_{\mathrm{SNeIa}}$. This results in $s_{00}<0$, which, in general, is required by physical processes such as vacuum gravitational Cherenkov radiation \cite{Kostelecky:2015dpa}, in binary-pulsars timing~\cite{Shao:2014oha,Shao:2014bfa}, and also Very-Long Baseline Interferometry (VLBI)~\cite{LePoncin-Lafitte:2016ocy}. To obtain an exact value for the timelike background field $s_{00}$, we employ Eq.~\eqref{H2}, which allows us to derive the following relationship:
\begin{subequations}
	\label{eq:delta-H-scenario-1}
	\begin{equation}
		\delta H =\sqrt{1+\frac {s_{00}}{2(1-s_{00} )}}-1\,,
		\label{H3}
	\end{equation}
	where
	\begin{equation}
		\delta H\equiv\frac{\Delta H}{H_{\mathrm{SNeIa}}}\,,\quad \Delta H\equiv H_{\mathrm{CMB}}-H_{\mathrm{SNeIa}}\,.
	\end{equation}
\end{subequations}
Now, by taking into account the recent reports on the values of the Hubble parameters $H_{\mathrm{CMB}}$ and $H_{\mathrm{SNeIa}}$ related to the early Universe and the current epoch, respectively, we obtain $-0.11<\delta H_{\mathrm{HT}}<-0.07$, meaning that $\delta H_{\mathrm{HT}}$ arising from the HT takes values from $-0.11$ to $-0.07$ at the $1\,\sigma$ confidence level. Besides, for the scenario at hand being able to account for the HT, $\delta H$ of Eq.~\eqref{eq:delta-H-scenario-1} should not exceed the value of the HT, i.e., $\delta H<\delta H_{\mathrm{HT}}$.

By considering the $1\,\sigma$ range for $\delta H_{\mathrm{HT}}$, we obtain upper bounds $s_{00}<-0.71$ and $s_{00}<-0.36$, respectively.
In other words, by adopting a conservative approach, if $s_{00}$ takes values smaller than $-0.36$ according to the HT, the extended cosmology of the first scenario in Sec.~\ref{sec:first-scenario} can be a potential candidate for describing early cosmology. However, despite the preference of negative values for $s_{00}$, by referring to the previous Refs.~\cite{Shao:2014oha,Shao:2014bfa,Kostelecky:2015dpa,LePoncin-Lafitte:2016ocy}, one infers that $s_{00}<-0.36$ has already been strictly ruled out.
Hence, by accepting the HT as a speculative criterion to investigate modified cosmologies of the early Universe, the extended scenario of Sec.~\ref{sec:first-scenario} is not supported. However, this statement becomes invalid when future measurements rule out the HT.

By turning to the second scenario of Sec.~\ref{sec:second-scenario}, from the modified first Friedmann equation of \eqref{eq:frdmod}, we have
\begin{subequations}
	\begin{equation}
		H_{\mathrm{CMB}}^2 =H^2_{\mathrm{SNeIa}}(1-\alpha a^2)^{-1}\,,
		\label{H4}
	\end{equation}
	which results in
	\begin{equation}
		\delta H=(1-\alpha a^2)^{-1/2}-1\,.
		\label{H5}
	\end{equation}
\end{subequations}
So the second scenario can explain the HT provided that the modification satisfies the upper bounds $\alpha a^2<-0.26$ and $\alpha a^2<-0.15$ corresponding to the boundary values $-0.11$ and $-0.07$, respectively, of the $1\,\sigma$ confidence interval for $\delta H_{\mathrm{HT}}$. Besides, here we must deal with the theoretical ranges on the scale factor in Eqs.~\eqref{eq:range-scale-factor-dust} and \eqref{eq:range-scale-factor-radiation} for dust and radiation, respectively. The second is automatically included in the first. By applying a bit of algebra to Eq.~\eqref{eq:range-scale-factor-radiation}, it can be re-expressed as
\begin{equation}
	\frac{1}{3}<\alpha a^2<1\,.
	\label{H7}
\end{equation}
A comparison of the latter to the above lower constraints $\alpha a^2$ tells us that they do not overlap. This means that the theoretical range on the model parameter is not consistent with the bounds derived from the HT. Thus, the HT is in conflict with the scenario of Sec.~\ref{sec:second-scenario}, too. It is also worthwhile to mention that the rather large values of the controlling coefficients determined previously do not contradict the use of the FLRW metric, since the coefficients $s_{00}$ and $\alpha$ neither depend on the spacetime coordinates nor do they directly give rise to anisotropies in local inertial frames.

Interestingly, according to several previous studies such as \cite{Krishnan:2021dyb,Krishnan:2021jmh,Aluri:2022hzs,Luongo:2021nqh}, deviations from the FLRW metric contribute to a resolution of the HT. In other words, these models, which deviate sufficiently from FLRW cosmology, can lower the sound horizon and raise the Hubble parameter, since they modify the time evolution of the early Universe. This occurs in a way that is consistent with data from baryon acoustic oscillations (see the review paper \cite{Abdalla:2022yfr} for more details).

Now, we are tempted to say that generic Lorentz and diffeomorphism violation in SME cosmologies, if so large enough, can be a source for deviations from FLRW cosmology. However, the very tight constraints compiled for various coefficients of background fields in the SME gravity sector (see the data tables~\cite{Kostelecky:2008ts}) eliminate such a possibility. Hence, this kind of symmetry violation seems to be too small to imply meaningful deviations from FLRW cosmology.


\section{Conclusions}
\label{sec:conclusions}

In this paper, our goal was to account for the Hubble crisis via cosmological models of the pre-CMB era. We focused on two cosmological settings based on particular modifications of GR parameterized by the gravitational SME. The first \cite{ONeal-Ault:2020ebv} was characterized by a single coefficient $s_{00}$ of the tensor-valued background field $s_{\mu\nu}$, whereas the second \cite{Reyes:2022dil} was governed by a diagonal choice of the purely spacelike subset $s^{ij}$ of the background field $s^{\mu\nu}$.

Both background fields were assumed to be nondynamical, i.e., they imply violations of diffeomorphism symmetry. Note also that the background field in the first (second) scenario has lower (upper) indices, which is one of the reasons why both theories lead to a vastly different cosmological evolution. As was shown in Ref.~\cite{Reyes:2022dil}, the second model is capable of describing an accelerated expansion of the Universe without considering any type of exotic matter component to do so.

In particular, our intention was to find out whether or not each of the two modified cosmologies was capable of accounting for the HT, which is a discrepancy in the experimental measurements of the Hubble parameter based on early cosmology and the current epoch, respectively. We found that none of the two scenarios governed by $s_{00}$ and $s^{ij}$, respectively, can explain the HT. This can be interpreted as a step towards showing the inability of new physics beyond GR, particularly inferred from the gravitational SME, to solve the problem of the HT. It is also very important to note that postulating the existence of any new physics based on the HT is merely speculative until the nature of this discrepancy in the value of the Hubble parameter is disclosed. As a result, the main message of our manuscript is to demonstrate how the HT can serve as a potential criterion for assessing modified cosmologies that rest upon the gravitational SME. Overall, this statement applies to refining all extended cosmologies beyond $\Lambda$CDM addressing the early Universe.

Generic SME coefficients may act as the source for deviations from the FLRW metric if they are large enough. It is tempting to resort to this property, especially when taking into account that deviations from the FLRW metric can contribute to solving the issue of the HT \cite{Krishnan:2021dyb,Krishnan:2021jmh,Aluri:2022hzs,Luongo:2021nqh}. However, two reasons rule out the idea of new physics beyond FLRW that may originate from symmetry violation in the SME cosmologies referred to in the present work. First, the SME coefficients are free of inhomogeneous and anisotropic properties. Second, the data tables \cite{Kostelecky:2008ts} report very tight constraints on coefficients of background fields in the gravitational SME.

One last statement refers to the possibility of field redefinitions that allow for moving coefficients between different sectors of the SME. For example, field redefinitions exist that remove $s_{\mu\nu}$ from the gravity sector such that the fermion sector coefficients $c_{\mu\nu}$ and the nonbirefringent photon sector coefficients $\tilde{\kappa}_{\mu\nu}\equiv (k_F)^{\alpha}_{\phantom{\alpha}\mu\alpha\nu}$ get shifted accordingly \cite{Bluhm:2019ato}. Thus, an alternative to studying the $s_{\mu\nu}$-related contributions in the Friedmann equations is to keep the standard dependence on the scale factor $a(t)$ intact and to include Lorentz violation in the matter sector, in particular, in terms of the $c_{\mu\nu}$ and $\tilde{\kappa}_{\mu\nu}$ coefficients previously referred to.

The data tables \cite{Kostelecky:2008ts} tell us that the sensitivities of the isotropic coefficients $c_{TT}^{(e,p,n)}$ for electrons, protons, and neutrons, respectively, are $10^{-21}$, $10^{-22}$, and $10^{-13}$. The single dimensionsless isotropic coefficient in the photon sector has been constrained to the level of $10^{-19}$. Therefore, we are tempted to conclude that isotropic Lorentz violation in the matter sector cannot account for the Hubble tension, as well, since the latter is a percentage effect. So we are also in a position to rule out such kinds of models as explanations of this discrepancy in the data.

However, we are not in a position to draw conclusions on models parameterized by alternative sets of coefficients such as the string-inspired model reviewed in the recent Ref.~\cite{Mavromatos:2022gtr}. The latter involves an axion background described by a purely timelike fermionic background field $b_{\mu}$, which gives rise to a matter-antimatter asymmetry in leptons.

\section*{Acknowledgments}
We are grateful to V.A.~Kosteleck\'{y} for his insightful reading and comments on this manuscript. We thank the anonymous referee for the helpful comments that improved the quality of the manuscript. M.Kh thanks Shiraz University Research Council. M.S. is indebted to FAPEMA Universal 00830/19, CNPq Produtividade 310076/2021-8, and CAPES/Finance Code 001.

\end{document}